\documentclass[twocolumn,showpacs,preprintnumbers,amsmath,amssymb]{revtex4}

\usepackage{graphicx}
\usepackage{dcolumn}
\usepackage{bm}

\begin{document}

\title{A Dirac Sea of Quantized Superluminal States in the Kerr-Newman Disk}
\author{Robert D. Bock\footnote{email: rdbock@gmail.com}}

\affiliation{XX}

\date{\today}

\begin{abstract}
We model the continuous material component of the Kerr-Newman disk as a discrete distribution of superluminal particles undergoing quantized orbital rotation with velocity $v/c = a/r$.  By matching the discrete distribution in the continuous limit with the Kerr disk densities, we derive the quantized energy and momentum states of the superluminal orbits.  We conclude that the Kerr-Newman disk may be modeled as a Dirac sea of these superluminal states and we identify possible excited states of the disk.
\end{abstract}

\maketitle
The Kerr-Newman metric \cite {kerr,newmanetal} is an exact, stationary, axially symmetric vacuum solution of the coupled Einstein-Maxwell field equations and describes the gravitational field of a mass endowed with angular momentum and charge.  Newman et al. \cite{newmanetal} originally obtained this metric from a complex coordinate transformation of the Reissner-Nordstrom metric, a procedure \cite{newmanjanis} that previously was shown to transform the Schwarzschild solution into the rotating Kerr metric.  The Kerr vacuum is often used to model black holes, however, it may be also used to model the exterior gravitational field of a rotating astrophysical mass, provided the solution is assumed to be valid outside some compact region of the object.

One may define a coordinate $r$ in the Kerr manifold, which can be defined geometrically as an affine parameter along either of the two congruences of principal null geodesics.  This coordinate functions as an asymptotic radial coordinate at spatial infinity.  At $r=0$, the Kerr geometry possesses an equatorial disk that is centered on the axis of symmetry.  In the absence of charge, this disk is intrinsically flat and of radius $|a|=\left| \frac{J_0}{m_0c} \right|$, where $m_0$ and $J_0$ are the gravitational mass and angular momentum observed at large $r$.  On the boundary of the disk there is a geometrical singularity.  In addition, the Kerr-Newman field exhibits the Dirac gyromagnetic ratio, twice that of a classical distribution.  

Israel \cite{israel} investigated the source of the Kerr-Newman field and derived the mass, angular momentum, and charge distributions on the disk using the theory of surface layers in general relativity.  He adopted the interpretation that the complete Kerr manifold is defined so that $r\geq 0$ everywhere and that the discontinuity in the normal derivative of the metric across the equatorial disk at $r=0$ is due to a layer of mass and charge.  Israel discovered that the material component of the disk rotates with superluminal velocity and that the charge rotates with subluminal velocity, both approaching the speed of light as $r\rightarrow |a|$.  He discovered that the disk has negative surface energy, charge, and angular momentum densitites, that diverge to $-\infty$ as one approaches the ring singularity.  He postulated that this is rectified at the singular ring by a positively infinite mass, angular momentum, and charge.  
 
Since classical densities are generally viewed as distributions possessing an underlying quantized structure, we examine Israel's distributions in the context of discrete orbital states.  In particular, we model the material component of the Kerr disk as superluminal particles with intrinsic spin undergoing orbital rotation, restricted by the de Broglie quantization condition  \cite{debroglie}.  We conclude that the material component of the Kerr disk may be modeled as a Dirac sea of superluminal particles.  We derive the quantized energy states and examine what restrictions quantum theory has on the occupancy at each level.  We also discuss excitations of the disk and implications for the mass spectrum of the fundamental particles.
 
In order to model the material component, we use the uncharged Kerr disk since the mass and charge densities rotate at different speeds.  In this case the disk is intrinsically flat and $r$ may be considered the radial coordinate.  Taking kinetic and gravitational potential energy into account, the ``effective'' energy density is \cite{israel}:
\begin{equation}
\label{energydensity}
\sigma_{eff}=-\frac{m_0c^2}{2\pi a^2}\frac{1}{\left(1 - \frac{r^2}{a^2} \right)^{3/2}}.
\end{equation}
The material component of the disk rotates with superluminal speed $v/c = a/r>1$ as measured by a stationary observer at infinity.  In addition, the annulus from $r$ to $r + dr$ contributes angular momentum:
\begin{equation}
\label{angularmomentumcontribution}
dJ = -\frac{m_0c}{2}\left(\frac{r}{a} \right)^3\frac{dr}{\left(1-\frac{r^2}{a^2} \right)^{3/2}}.
\end{equation}
Israel noted that both the energy and angular momentum diverge as $r\rightarrow |a|$, therefore, he concluded that the singular ring must contribute an infinite positive mass and angular momentum, in order to yield the finite net observed values $m_0$ and $J_0$.

We model the material density as quantized superluminal particles  \cite{feinberg} of (imaginary) mass $\mu$ with angular velocity $v/c = a/r > 1$.  We write the total energy of these states as
\begin{equation}
\label{superluminalenergy}
E=\frac{\mu c^2}{\sqrt{\frac{v^2}{c^2}-1}} + V(r),
\end{equation}
where $V(r)$ is a potential that binds the particles to the singular ring.  The momentum of each state is:
\begin{equation}
\label{superluminalmomentum}
|p|=\frac{\mu v}{\sqrt{\frac{v^2}{c^2}-1}}.
\end{equation}
Note that the four-momentum of the free superluminal state is a space-like vector and $c^2p^2-E^2 = \mu^2 c^4$.  We assume that the particles are bound to the singular ring by the potential V(r) and obey Fermi-Dirac statistics so that they are supported in their orbits by degeneracy pressure.  In other words, we assume that the centripetal acceleration ($F_c$), binding force ($F_b$), and supporting degeneracy force ($F_d$), bind the particles such that:
\begin{equation}
\label{forces}
F_d = F_c + F_b,
\end{equation}
with $v/c = a/r > 1$.  Below we will examine this assumption more carefully. 

It can be shown (see Appendix) that the de Broglie \cite{debroglie} quantization condition can be derived for the superluminal states.  Therefore, we consider the following quantization condition:
\begin{equation}
\label{orbitalquantization}
\frac{\mu v r}{\sqrt{\frac{v^2}{c^2} - 1}}=n\hbar,
\end{equation}
where $v/c = a/r$.
This gives the following quantized radii:
\begin{equation}
\label{quantizedradii}
\frac{r}{a}=\frac{1}{\sqrt{1+\left(\frac{ a/l }{n } \right)^2}},
\end{equation}
where $l\equiv \frac{\hbar}{\mu c }$.  In the limit that $\frac{a}{l}\rightarrow \infty$ Equation (\ref{quantizedradii}) yields a continuum of states.  We demand that when this continuum of states is filled it reproduces the energy and angular momentum densities of the Kerr disk (Equations (\ref{energydensity}) and (\ref{angularmomentumcontribution})), which we write as:
\begin{equation}
\label{dEdr}
\frac{dE}{dr}=(2\pi r)\sigma_{eff}=-\frac{m_0c^2}{ a^2}\frac{r}{\left(1 - \frac{r^2}{a^2} \right)^{3/2}}
\end{equation}
and
\begin{equation}
\label{dJdr}
\frac{dJ}{dr}=-\frac{m_0c}{2}\left(\frac{r}{a} \right)^3\frac{1}{\left(1-\frac{r^2}{a^2} \right)^{3/2}}.
\end{equation}
Using these Kerr distributions, we may calculate the total energy and angular momentum present at each quantum number $n$ by noting:
\begin{eqnarray}
\label{chain}
\frac{dE}{dr} &=& \frac{dE}{dn}\frac{dn}{dr} \\ \nonumber
\frac{dJ}{dr} &=& \frac{dJ}{dn}\frac{dn}{dr}.
\end{eqnarray}
Using Equation (\ref{quantizedradii}) we write:
\begin{equation}
\label{index}
n = \left( \frac{a}{l} \right) \frac{1}{\sqrt{\frac{a^2}{r^2}-1}}
\end{equation}
so that:
\begin{equation}
\label{dndr}
\frac{dn}{dr} = l^{-1}\frac{1}{\left(1-\frac{r^2}{a^2} \right)^{3/2}}.
\end{equation}
Therefore, using Equations (\ref{dEdr}), (\ref{dJdr}), (\ref{chain}), and (\ref{dndr}), we obtain:
\begin{equation}
\label{dEdn}
\frac{dE}{dn}=-\left (\frac{l}{a} \right)\frac{m_0c^2}{\sqrt{1+\left(\frac{ a/l }{n} \right)^2}}
\end{equation}
and 
\begin{equation}
\label{dJdn}
\frac{dJ}{dn} = -\frac{J_0}{2} \left(\frac{l}{a} \right) \left[\frac{n^3}{\left(n^2 + \left(\frac{a}{l}\right)^2\right)^{3/2}}  \right]. 
\end{equation}
Integrating the above equations gives:
\begin{equation}
\label{E}
E=-m_0c^2 \left(\frac{l}{a} \right)\sqrt{n^2 + \left(\frac{a}{l}\right)^2} + E^{(0)}
\end{equation}
and
\begin{equation}
\label{J}
J = -\frac{J_0}{2} \left(\frac{l}{a} \right) \left[\frac{n^2 + 2\left(\frac{a}{l}\right)^2}{\sqrt{n^2 + \left(\frac{a}{l}\right)^2}}  \right] + J^{ (0) },
\end{equation}
where $E^{(0)}$ and $J^{(0)}$ are constants of integration.  Note that $J$ gives the total angular momentum (orbital + intrinsic spin) present at orbital $n$.  In order for the continuum of states to be filled throughout the disk, we must assume that the superluminal particles obey Fermi-Dirac statistics, otherwise, only the lowest energy state ($n=\infty$) would be filled.  It does not necessarily follow, however, that they are spin-1/2 particles \cite{feinberg}, although it is possible.  We can solve for the constants of integration by evaluating the above at $n = 0$, since the velocity $v=\infty$ is allowed in the superluminal regime.  Noting that at $n=0$ the particle energy is $E=V(r=0)$ and the orbital angular momentum vanishes, we obtain:
\begin{equation}
\label{E0}
E^{(0)} = m_0 c^2 + V(r=0)
\end{equation}
and
\begin{equation}
\label{J0}
J^{(0)} = J_0 + s_{z}(n=0),
\end{equation} 
where $s_{z}(n=0)$ is the projection of the total intrinsic spin of the superluminal particles in the state $n=0$ along the axis of symmetry.  

Since the magnitude of the orbital angular momentum increases by integral values of $\hbar$ as $n$ increases, one must assume that the total intrinsic spin accounts for the nonlinear functional change in $J$.  Therefore, we write:
\begin{equation}
\label{J2}
J\ = -fn\hbar + s_z(n),
\end{equation}
where $f$ is the number of particles in each state and 
\begin{eqnarray}
\label{sz}
s_z(n) = -\frac{J_0}{2} \left(\frac{l}{a} \right) \left[\frac{n^2 + 2\left(\frac{a}{l}\right)^2}{\sqrt{n^2 + \left(\frac{a}{l}\right)^2}}  \right]  \nonumber \\ + J_0 + s_{z}(n=0) + fn\hbar
\end{eqnarray}
is the total intrinsic spin along the axis of symmetry of the superluminal particles as a function of $n$.  In the continuous limit ($\frac{a}{l}\rightarrow \infty$) with low $n$, $\frac{a}{l}\gg n$, Equations (\ref{E}) and (\ref{sz}) become:
\begin{equation}
\label{Esmalln}
E\simeq  -\frac{m_0 c^2}{2}\left(\frac{n}{a/l} \right)^2  + V(r=0)
\end{equation}
and
\begin{equation}
\label{szcontinuous}
s_z(n) \simeq -\left[\frac{J_0}{4}\left(\frac{n}{\left(a/l \right)}  \right)^4 - fn\hbar \right] + s_{z}(n=0).
\end{equation}
The opposite limit $\frac{a}{l}\ll n$ can be realized either in the limit $\frac{a}{l}\gg 1$ as $n\rightarrow \infty$ or in the limit $\frac{a}{l}\ll 1$ so that $\frac{a}{l}\ll n$ for all $n$.  In this case Equations (\ref{E}) and (\ref{sz}) become:
\begin{eqnarray}
\label{Elargen}
E\simeq -m_0c^2 \frac{n}{\left(\frac{a}{l} \right)}\left(1 + \frac{1}{2}\left(\frac{a/l}{n} \right)^2  \right)\nonumber \\ + m_0 c^2 + V(r=0)
\end{eqnarray}
and
\begin{eqnarray}
\label{Jlargen}
s_z(n)\simeq -\frac{J_0}{2}\frac{n}{\left(\frac{a}{l} \right)} \left[ 1 + \frac{3}{2}\left(\frac{a/l}{n} \right)^2 - \left(\frac{a/l}{n} \right)^4 \right] \nonumber \\+ fn\hbar  + J_0 + s_{z}(n=0).
\end{eqnarray}
Equations (\ref{E}) and (\ref{J}) were derived in the limit $\frac{a}{l}\rightarrow \infty$ and may not be valid outside of this continuous limit.  However, since the states are only a function of radius, it is possible that the same energy and momentum states will be realized throughout the entire range of $\frac{a}{l}$.  Thus, we assume that Equations (\ref{E}) and (\ref{J}) are valid for all $\frac{a}{l}$, understanding that this may not necessarily be the case outside of the limit in which they were derived.  

Note that it may be impossible to populate states above some maximum orbit, $n_{max}$, depending on the intrinsic spin of each superluminal particle, the number of superluminal particles per orbit, and the angular momentum of the Kerr metric observed at large $r$, because the magnitude of $s_z$ may unbounded and there will be a limit to the number of particles allowed at each orbit.  In the limit $\frac{a}{l}\ll n$ this occurs if
\begin{equation}
\label{sunbounded}
J_0>2f\hbar \left(\frac{a}{l}\right).
\end{equation}

In general, the states between a minimum orbit, $n_{min}$, and a maximum orbit, $n_{max}$, will be occupied.  $n_{min}$ defines the highest occupied energy level and $n_{max}$ represents the lowest occupied energy level.  The Kerr disk is realized in the continuous limit ($\frac{a}{l}\rightarrow \infty$) with $n_{min}=0$ and $n_{max}=\infty$, so that the entire disk is a  Dirac sea of superluminal states.  If the magnitude of $s_z$ is unbounded, for example, according to (\ref{sunbounded}), then the disk may still be completely filled, however, one must then introduce an infinite number of non-identical superluminal particles as $n \rightarrow \infty$. We will not examine the resulting gravitational field here for a partially-filled Kerr disk, however, it would be of interest to see whether or not some of the problems associated with the Kerr-Newman metric, such as closed timelike curves and naked singularities, are resolved by restricting the occupancy to a finite orbital range.   

Excitations will occur for changes in the occupancy at a given orbital state.  For example, a particle at state $n$ may be ejected from the disk. If no other particles enter the disk during this transition then a ``hole'' will exist at the previously occupied orbit.  This hole, which represents an absence of negative mass and negative angular momentum, will manifest itself as an increase in the gravitational mass $m_0$ and angular momentum $J_0$ of the disk as measured at spatial infinity.  Similarly, an empty or partially filled orbital state may be filled by capturing an external superluminal particle or from transitions within the disk.  In addition, if an empty or partially filled orbit exists on the disk near $n_{min}$, then excited states of the disk exist for the transition $n\rightarrow n -1$, which will increase both $m_0$ and $J_0$ of the disk.  

Note that angular momentum (cf. Equation (\ref{J})) may severely restrict such processes.  If we demand that Equation (\ref{J}) holds before and after any changes in the disk occupancy, then the transitions must satisfy the selection rule
\begin{equation}
\label{deltaj}
\Delta J = 0.
\end{equation}
Consequently, many transitions will either be forbidden or will be accompanied by a realignment of the intrinsic spins in order to satisfy Equation (\ref{deltaj}).  

A filled orbital state at $n_{max}$ may capture a non-identical superluminal particle; this may or may not be accompanied by the ejection of a superluminal particle, depending on the selection rule (\ref{deltaj}).  Consider the following scattering process for a Kerr metric with gravitational mass $m_0$ and angular momentum $J_0$ that is  filled up to an orbit $n_{max}$.  A superluminal particle of mass $\mu_a$ and spin $s_a$ at orbit $n_{max}$ may be replaced by another superluminal particle of mass $\mu_b$ and spin $s_b$.  In terms of a Feynman diagram, this will appear as an interaction betwen an incoming particle ($m_0,J_0$) with the incoming superluminal particle ($\mu_b,s_b$), resulting in an outgoing particle of different mass and angular momentum ($m_0^*, J_0^*$) and an outgoing superluminal particle ($\mu_a, s_a$).  If the selection rule (\ref{deltaj}) is satisfied during this interaction then this scenario resembles certain weak interactions, since it has been speculated that neutrinos travel at superluminal speeds \cite{chodosetal}.  For example, in the example above, an electron ($m_0,J_0$) and a muon neutrino ($\mu_b, s_b$) may be considered the incoming particles, and a muon ($m_0^*, J_0^*$) and an electron neutrino ($\mu_a, s_a$) may be considered the outgoing particles, with $J_0 = J_0^*$.  

This suggests that the above Dirac sea model of the Kerr-Newman disk \footnote {Note that the uncharged Kerr disk may be a valid approximation for the electron since $a>>q$, where $q^2 = Ge^2/c^4$ and $e$ is the electron charge.} may be used to model the electron; furthermore, the muon and tau may be identified as excited states of the electron.  Note that the appropriate limit for the electron is $\frac{a}{l}\ll n$, since the neutrino masses are believed to be much smaller than the electron mass.  Of course, there are well-known problems with identifying the Kerr-Newman field with the gravitational field of an electron because the corresponding gravitational field  is a naked singularity and, moreover, it possesses closed timelike curves.  However, it was suggested above that the disk may not be completely filled because of angular momentum considerations and the resulting gravitational field may differ from the Kerr-Newman metric, possibly resolving such issues.  In fact, if one demands that Equation (\ref{Jlargen}) is satisfied for all orbital states, then only the $n=0$ state can be populated, since the spin of the neutrino is not sufficient to accommodate the higher ($n>0$) orbital states.  Thus, the electron may be viewed as the opposite limit of the Kerr disk, being a singular ring with a single superluminal state (electron neutrino) bound at $n=0$, with all other states unoccupied.  In addition, one may identify the muon and tau as excited states of the electron.  These excited states may be realized by the superluminal replacement process described above (an incoming muon neutrino or tau neutrino and an exiting electron neutrino).  The only excited states that exist correspond to the number of superluminal particles that can interchange with the electron neutrino, which must be identified with the lowest energy state when bound to the singular ring.  Thus, it is the number of neutrino types that restricts the number of members in the lepton family.  It follows that the electron-muon and electron-tau energy differences correspond to the differences in binding energy of the bound superluminal states. 

The above model also provides insight into the structure of the weak bosons and quarks.  Examination of the Feynman diagrams for weak interactions suggests that a fundamental lepton consists of the associated neutrino bound to a charged weak boson at the $n=0$ state.  For example, the electron consists of an electron neutrino bound to a $W^-$ boson.  It follows that the charged weak boson consists of the charged component of the disk and the singular ring.  Using the example stated above with an incoming electron and muon neutrino and an outgoing muon and electron neutrino, the $W^-$ boson can be viewed as transferring the singular ring and charged component of the disk from the electron to the muon neutrino, resulting in an outgong muon.  Similar considerations suggest that the neutral $Z^0$ boson consists of the angular momentum component of the singular ring.  In addition, consider an incoming electron and up quark and an outgoing electron neutrino and down quark, mediated by the $W^-$ boson.  Again, the $W^-$ boson can be viewed as transferring the singular ring and charged component of the disk to the up quark, resulting in a down quark.  Thus, the quarks may be considered charged excitations of the disk.  This will be discussed in more detail in a following communication.

In closing, we discuss the binding potential $V(r)$.  Of the four fundamental interactions, only the weak and gravitational forces may be considered to be the binding potential, since the neutrinos do not respond to the strong and electromagnetic forces.  However, it does not seem possible to satisfy the velocity condition $v/c = a/r$ with either gravity or the weak force.  More research in this direction is required.  If such efforts ultimately fail, then one will be led to postulate a new force to bind the superluminal particles to the singular ring.

\section{\label{sec:appendix}Appendix}
Following de Broglie's original analysis \cite{debroglie}, we derive the orbital quantization condition (\ref{orbitalquantization}) of the superluminal states.  The energy and momentum of a free superluminal particle are given by Equation (\ref{superluminalenergy}) with $V(r) = 0$ and Equation (\ref{superluminalmomentum}), respectively.  Note that $|pc|>E$ and the velocity may still be defined by $v/c = |p|/E>1$.  Unlike their subluminal counterparts, the range of velocity is $\infty\ge|v|> c$, with $v=\infty$ permitted.  In addition, the corresponding ranges of energy and momentum are:
\begin{eqnarray}
\label{Epranges}
0\le &E&<\infty \\ \nonumber
\mu c \le &|p|& <\infty.
\end{eqnarray}
Note that energy and momentum reverse roles in the superluminal regime, with the lower bound of energy vanishing and the lower bound of momentum remaining finite.  

In his original analysis, de Broglie postulated a periodic phenomenon $\exp (i \nu_0 t)$ in the rest frame ($v=0$) of the particle with frequency defined by:
\begin{equation}
\label{debrogliefrequency}
h\nu_0 = m_0 c^2.
\end{equation}
He noted that this relationship remained relativistically invariant if $\nu_0$ was interpreted as the frequency of a wave.  Consequently, the relationship $p = h/\lambda$ followed from this interpretation.  

We define the superluminal ``rest'' frame of the particle as $v_\infty=\infty$.  In this frame, the energy vanishes, and therefore Equation (\ref{debrogliefrequency}) must be replaced by a relationship with the non-vanishing ``rest'' momentum.  Following de Broglie, we postulate a periodic phenomenon $\exp (i k_0 x)$ in this rest frame with
\begin{equation}
\label{k0}
\hbar k_0 \equiv h/\lambda = \mu c .
\end{equation}
Under a Lorentz boost of velocity $u$, this becomes:
\begin{equation}
\label{transformedframe}
\exp  \left[i k_0\left(\frac{x - u t}{\sqrt{1 - \frac{u^2}{c^2}}}\right) \right].
\end{equation}
Therefore, under the Lorentz boost, the periodic phenomenon becomes:
\begin{equation}
\label{ktransformed}
\hbar k = \frac{\hbar k_0}{\sqrt{1 - \frac{u^2}{c^2}}}.
\end{equation}
Hence, Equation (\ref{k0}) is relativistically invariant if $k$ is associated with a wave of frequency:
\begin{equation}
\label{nutransformed}
h\nu = \mu c^2\frac{u/c}{\sqrt{1 - \frac{u^2}{c^2}}}.
\end{equation}
A Lorentz transformation with velocity $u$ from the frame $v_\infty$ boosts the particle to the speed $v$ given by:
\begin{equation}
\label{boostedvelocity}
v = \frac{v_\infty - u}{\left(1 - \frac{u v_\infty}{c^2}   \right)}.
\end{equation}
In the limit $v_\infty \rightarrow \infty$, this gives $u\rightarrow -c^2 / v$.  Substituting this into Equation (\ref{ktransformed}) and using Equation (\ref{k0}) yields the superluminal de Broglie relationship:
\begin{equation}
\label{pequalshoverlambda}
\frac{h}{\lambda} = \frac{\mu v}{\sqrt{\frac{v^2}{c^2}-1}} \equiv p.
\end{equation}
Similarly, Equation (\ref{nutransformed}) becomes:
\begin{equation}
\label{nutransformed2}
h\nu = \frac{\mu c^2}{\sqrt{\frac{v^2}{c^2}-1}}.
\end{equation}
Demanding that an integral number of wavelengths fits around the circumference at radius $r$ gives the quantization condition (\ref{orbitalquantization}).
 
\bibliography{main}

\begin{thebibliography}{7}
\expandafter\ifx\csname natexlab\endcsname\relax\def\natexlab#1{#1}\fi
\expandafter\ifx\csname bibnamefont\endcsname\relax
  \def\bibnamefont#1{#1}\fi
\expandafter\ifx\csname bibfnamefont\endcsname\relax
  \def\bibfnamefont#1{#1}\fi
\expandafter\ifx\csname citenamefont\endcsname\relax
  \def\citenamefont#1{#1}\fi
\expandafter\ifx\csname url\endcsname\relax
  \def\url#1{\texttt{#1}}\fi
\expandafter\ifx\csname urlprefix\endcsname\relax\def\urlprefix{URL }\fi
\providecommand{\bibinfo}[2]{#2}
\providecommand{\eprint}[2][]{\url{#2}}

\bibitem[{\citenamefont{Kerr}(1963)}]{kerr}
\bibinfo{author}{\bibfnamefont{R.~P.} \bibnamefont{Kerr}},
  \bibinfo{journal}{Phys. Rev. Lett.} \textbf{\bibinfo{volume}{11}},
  \bibinfo{pages}{238} (\bibinfo{year}{1963}).

\bibitem[{\citenamefont{Newman et~al.}(1965)\citenamefont{Newman, Couch,
  Chinnapared, Exton, Prakash, and Torrence}}]{newmanetal}
\bibinfo{author}{\bibfnamefont{E.~T.} \bibnamefont{Newman}},
  \bibinfo{author}{\bibfnamefont{E.}~\bibnamefont{Couch}},
  \bibinfo{author}{\bibfnamefont{R.}~\bibnamefont{Chinnapared}},
  \bibinfo{author}{\bibfnamefont{A.}~\bibnamefont{Exton}},
  \bibinfo{author}{\bibfnamefont{A.}~\bibnamefont{Prakash}}, \bibnamefont{and}
  \bibinfo{author}{\bibfnamefont{R.}~\bibnamefont{Torrence}},
  \bibinfo{journal}{J. Math. Phys.} \textbf{\bibinfo{volume}{6}},
  \bibinfo{pages}{918} (\bibinfo{year}{1965}).

\bibitem[{\citenamefont{Newman and Janis}(1965)}]{newmanjanis}
\bibinfo{author}{\bibfnamefont{E.~T.} \bibnamefont{Newman}} \bibnamefont{and}
  \bibinfo{author}{\bibfnamefont{A.~I.} \bibnamefont{Janis}},
  \bibinfo{journal}{Journal of Mathematical Physics}
  \textbf{\bibinfo{volume}{6}}, \bibinfo{pages}{915} (\bibinfo{year}{1965}).

\bibitem[{\citenamefont{Israel}(1970)}]{israel}
\bibinfo{author}{\bibfnamefont{W.}~\bibnamefont{Israel}},
  \bibinfo{journal}{Phys. Rev. D} \textbf{\bibinfo{volume}{2}},
  \bibinfo{pages}{641} (\bibinfo{year}{1970}).

\bibitem[{\citenamefont{de~Broglie}(1923)}]{debroglie}
\bibinfo{author}{\bibfnamefont{L.}~\bibnamefont{de~Broglie}},
  \bibinfo{journal}{C. R. Acad. Sci.} \textbf{\bibinfo{volume}{177}},
  \bibinfo{pages}{507} (\bibinfo{year}{1923}).

\bibitem[{\citenamefont{Feinberg}(1967)}]{feinberg}
\bibinfo{author}{\bibfnamefont{G.}~\bibnamefont{Feinberg}},
  \bibinfo{journal}{Physical Review} \textbf{\bibinfo{volume}{159}},
  \bibinfo{pages}{1089} (\bibinfo{year}{1967}).

\bibitem[{\citenamefont{Hauser and Kostelecky}(1985)}]{chodosetal}
\bibinfo{author}{\bibfnamefont{A.~C. A.~I.} \bibnamefont{Hauser}}
  \bibnamefont{and} \bibinfo{author}{\bibfnamefont{V.~A.}
  \bibnamefont{Kostelecky}}, \bibinfo{journal}{Phys. Lett.}
  \textbf{\bibinfo{volume}{B150}}, \bibinfo{pages}{431} (\bibinfo{year}{1985}).

\end{thebibliography}
\end{document}